\documentclass[english,showpacs,floatfix,preprint,12pt]{revtex4}

\usepackage[dvips]{graphicx}
\usepackage{longtable}
\usepackage{amsmath,amssymb}
\usepackage{dcolumn}
\usepackage{latexsym}
\usepackage{babel}
\usepackage[latin1]{inputenc}

\newcommand{\be}{\begin{equation}}
\newcommand{\ee}{\end{equation}}

\begin{document}
\title{Le Ch\^{a}telier-Braun principle in cosmological physics}
\author{Diego Pav\'{o}n\footnote{E-mail address: diego.pavon@uab.es}}
\affiliation{Department of Physics,  Autonomous University of
Barcelona, 08193 Bellaterra}
\author{Bin Wang\footnote{E-mail address:
wangb@fudan.edu.cn}} \affiliation{Department of Physics, Fudan
University, 200433 Shanghai}

\begin{abstract}
Assuming that dark energy may be treated as a fluid with a well
defined temperature, close to equilibrium, we argue that if
nowadays there is a transfer of energy between dark energy and
dark matter, it must be such that the latter gains energy from the
former and not the other way around.
\end{abstract}


\maketitle

Recently, there have been some proposals in the literature to the
effect that the dark energy and dark matter should not conserve
separately but interact with each other \cite{interact}. As is
well known, the interaction may substantially alleviate the
coincidence problem \cite{substantially}, explain why
observationally the equation of state parameter of dark energy,
$w_{x} = p_{x}/\rho_{x} < -1/3$, may seem of phantom type (i.e.,
lower than $-1$) \cite{bin3,das}, and account for the age of the
quasar APM0879+5255  at redshift $z=3.91$ \cite{bin}. However,
there is not consensus about whether the overall transfer of
energy should go from dark energy to dark matter
\cite{substantially,bin3,bin2,bin} or viceversa \cite{das},
\cite{viceversa}. In this short communication, we assume that both
components (dark matter and dark energy) are amenable to a
thermo-fluid description and resort to the second law of
thermodynamics to discern the sense in which the transfer of
energy proceeds. Obviously, this approach does not apply if dark
energy is a scalar field in a pure quantum state since its entropy
vanishes and no temperature can be defined. However, this is not
the more general case. One may consider dark energy as the
effective manifestation of a mixture of scalar fields. Generally,
the mixture would not be in a pure quantum state whereby it would
be entitled to an entropy and a global temperature.

If these two components conserved separately in an expanding
Friedmann-Lema\^{i}tre-Robertson-Walker universe we would have
\\
\begin{eqnarray}\label{consv1a}
\dot{\rho}_m&+&3H(\rho_m+ p_{m})=0 \, , \\
\label{consv1b}
\dot{\rho}_{x}&+&3H(1+w_{x})\rho_{x}=0\, ,
\end{eqnarray}
\\
where the equation of state of dark matter can be approximately
written in parametric form as \cite{degroot}
\\
\begin{equation}\label{matter-pressure}
\rho_{m} = n_{m} \, M + \textstyle{3\over{2}}\,  n_{m} \, T_{m}\,
,\quad \quad p_{m} = n_{m} \, T_{m} \qquad  (k_{B} = 1)
\end{equation}
\\
so long as $T_{m} \ll M$. As a consequence, $\rho_{m} \sim a^{-3}$
and $\rho_{x} \propto \exp \int{-3(1+w_{x})\, da/a}$.

The temperatures dependence on the scale factor
\\
\begin{equation}\label{tevol1}
T_m \propto a^{-2} \, , \qquad \qquad T_x \propto \exp \int{-3
w_{x}\, da/a}
\end{equation}
\\
follow from integrating the evolution equation, $\dot{T}/T = -3 H
(\partial p/\partial \rho)_{n}$. The latter is straightforwardly
derived from Gibbs' equation, $T dS = d(\rho/n) + p \, d(1/n)$,
and the condition for $dS$ to be a differential expression,
$\partial^{2}S/(\partial T\,
\partial n)=
\partial^{2}S/(\partial n\, \partial T)$.

Equations (\ref{tevol1}) suggest that currently $T_{m} \ll T_{x}
$, and viceversa at very early times (modulo, $T_m \ll M$ still at
that times).

When both components interact, Eqs. (\ref{consv1a}) and
(\ref{consv1b}) generalize to
\\
\begin{eqnarray}\label{consv2a}
\dot{\rho}_m&+&3H(\rho_m+ p_{m})= Q \, , \\
\label{convs2b}
\dot{\rho}_{x}&+&3H(1+w_{x})\rho_{x}= -Q \, ,
\end{eqnarray}
\\
respectively, where $Q$ denotes the interaction term. (We note in
passing that the overall energy density $\rho_{m} +\rho_{x}$ is
conserved). Obviously, for $Q >0$ the energy proceeds from dark
energy to dark matter.

Assuming
\\
\begin{equation}\label{Q1}
 Q = 3H \lambda \, \rho_{x} \, ,
\end{equation}
with $\lambda$ a small constant, one follows $\dot{\rho}_{x} +
3H(1+w_{x}+ \lambda)\,  \rho_{x} = 0$. Thereby, $\dot{T}_{x}/T_{x}
= -3 (w_{x}+\lambda) \, \dot{a}/a$ and
\\
\begin{equation}\label{tx1}
T_{x} \propto \exp \int{-3\,(w_{x}+\lambda)\, \frac{da}{a}} \, .
\end{equation}
\\
Consequently, if $\lambda > 0$ (i.e., $Q > 0$), $T_{x}$ will
increase more slowly as the Universe expands than in the absence
of interaction and correspondingly $T_{m}$ will also decrease more
slowly. This is fully  consistent with the second law of
thermodynamics. The latter implies that when two systems that are
not in mutual equilibrium (thermal or otherwise) interact with one
another, the interaction tends to drive the systems to a final
common equilibrium. If they are left to themselves, they will
eventually reach equilibrium. In the case contemplated here the
equilibrium is never achieved because the expansion of the
Universe (that can be viewed as  an external agent) acts in the
opposite sense. If $\lambda$ were negative ($Q <0$), the
temperature difference would augment, something at variance with
the second law. Clearly, our conclusion critically rests on the
validity of Eq. (\ref{Q1}). Next, we will argue that this
expression approximately holds, at least piecewise.

In view of any of the two Eqs. (\ref{consv2a}) and
(\ref{convs2b}), $Q$ must be a function of the energy densities
multiplied by a quantity with units of inverse of time. For the
latter the obvious choice is the Hubble factor $H$; thus, $Q = Q(
H \rho_{m}\, , H \rho_{x})$. By power law expanding $Q$ and
retaining just the first term we may write $Q \simeq \lambda_{m}\,
H \rho_{m} + \lambda_{x}\, H \rho_{x}$. In the scaling regime,
i.e., when the ratio $\rho_{m}/\rho_{x}$ stays constant, Eq.
(\ref{Q1}) is readily recovered. Outside this particular regime
-at any rate, not at very early times- one expects that the said
ratio varies slowly (i.e., not much faster than the scale factor
$a(t)$) whence it might be considered piecewise constant. Thus, we
recover again Eq. (\ref{Q1}) though this time $\lambda$ is
constant only piecewise.

The foregoing scenario can be also understood in the light of the
Le Ch\^{a}telier-Braun principle: When a system is perturbed out
of its equilibrium state it reacts trying to restore it or achieve
a new one \cite{lechatelier}. At sufficiently early times $T_m >
T_x$, and the Universe expansion rapidly drives both systems to a
common -equilibrium- temperature $T^{(eq)}$ at say, $a =
a^{(eq)}$. However, subsequently $a > a^{(eq)}$ and the thermal
equilibrium is lost, $T_m < T^{(eq)} < T_x$.  In our case, the
answer of the system to the  equilibrium loss is a continuous
transfer of energy from dark energy to dark matter. Whereas this
does not bring the system to any equilibrium state it certainly
slows down the rate at which it moves away from equilibrium.
Notice that when $a < a^{(eq)}$ the system is approaching
equilibrium thereby it ``sees" no reason for any energy transfer.

We now turn to a very short reasoning, independent of expression
(\ref{Q1}), leading to $Q > 0$ as well. Following Zimdahl
\cite{winfried} the entropy production associated to our two
interacting fluids (modulo their chemical potentials vanish)
reads\footnote{See Eq. (24) of Ref. \cite{winfried}.}
\\
\begin{equation}\label{entropyprod}
S^{a}_{\, m;a} +  S^{a}_{\, x;a} = \left( \frac{1}{T_{m}}-
\frac{1}{T_{x}}\right)\, Q \,
\end{equation}
\\
(see also \cite{mauricio}), where $S^{a}_{\; i}$ -with $i = m, x$-
stand for the entropy flux of each fluid. From the second law,
$S^{a}_{\, m;a} + S^{a}_{\, x;a} \geq 0 $, and the fact that
nowadays $T_{m} < T_{x}$ is expected, the desired result follows.
At this point, a remark is in order. In the special -and
oftentimes- assumed case $w_{x} =$ constant the sound speed of
dark energy, $c_{s,x}$, is not longer given by $\, c_{s,x}^{2} =
\dot{p}_{x}/\dot{\rho}_{x}$ -see however, \cite{erickson}- whereby
the dark energy does not behave adiabatically and a further term
should be added to the right hand side of Eq. (\ref{entropyprod}).
This difficulty can be avoided in the more general case $w_{x}
\neq$ constant.

In spite of this reassuring outcome, owing to the fact that the
lower $\rho_{x}$, the larger $T_{x}$ the thermo-fluid description
might look suspicious. However, it is worth recalling that this is
very often the case for systems in which gravity plays a leading
role (as in Schwarzschild black holes). At any rate, it is
reasonable to expect that the said description breaks down at some
point, both when $a \ll 1$ and when $a \gg 1$. It must be added
that very likely a clear-cut answer to the validity of the
thermo-fluid approach will not be available until the nature of
dark energy gets elucidated. In the meantime the best we can do is
to explore all possible venues connected with this ingredient of
the cosmic budget.

Altogether, our analysis seems to indicate that, so long as dark
energy is amenable to a fluid description with a well defined
temperature not far from equilibrium, the overall energy transfer
should go from dark energy to dark matter if the second law of
thermodynamics and Le Ch\^{a}telier- Braun principle are to be
fulfilled. Interestingly enough, the result $Q > 0$ guarantees
that the ratio $\rho_{m}/\rho_{x}$ asymptotically tends to a
constant \cite{substantially}, \cite{luis}, thus greatly
alleviating the coincidence problem.

\acknowledgments{ The authors are grateful to the referee for
constructive comments. This work was partially supported by the
NNSF of China, Shanghai Education Commission, Science and
Technology Commission; the Spanish Ministry of Education and
Science under Grant FIS 2006-12296-C02-01, and the ``Direcci\'{o}
General de Recerca de Catalunya" under Grant 2005 SGR 000 87.}


\end{document}